# Microwave Q-band oscillator at 49GHz for broadband frequency conversion based on a Kerr optical soliton crystal micro-comb

Xingyuan Xu[1], Jiayang Wu[1], Mengxi Tan[1], Thach G. Nguyen[2], Sai T. Chu[3], Brent E. Little[4],
Roberto Morandotti[5,6,7], Arnan Mitchell[2], and David J. Moss[1]

[1]Centre for Micro-Photonics, Swinburne University of Technology, Hawthorn, VIC 3122, Australia. (Corresponding e-mail: dmoss@swin.edu.au).

[2]RMIT University, Melbourne, VIC 3001, Australia.

[3]Department of Physics and Material Science, City University of Hong Kong, Tat Chee Avenue, Hong Kong, China.

[4] State Key Laboratory of Transient Optics and Photonics, Xi'an Institute of Optics and Precision Mechanics, Chinese Academy of Science, Xi'an, China.

[5]INSR-Énergie, Matériaux et Télécommunications, 1650 Boulevard Lionel-Boulet, Varennes, Québec, J3X 1S2, Canada

[6] Visiting professor, ITMO University, St. Petersburg, Russia

[7] Visiting professor, Institute of Fundamental and Frontier Sciences, University of Electronic Science and Technology of China, Chengdu 610054, China.

**We report a broadband microwave frequency converter based on a coherent Kerr optical micro-comb generated by an integrated micro-ring resonator. The coherent micro-comb displays features that are consistent with soliton crystal dynamics with an FSR of 48.9-GHz. We use this to demonstrate a high-performance millimeter-wave local oscillator at 48.9-GHz in the Q-band for microwave frequency conversion. We experimentally verify the microwave performance up to 40 GHz, achieving a ratio of −6.8 dB between output RF power and IF power and a spurious suppression ratio of > 43.5 dB. The experimental results show good agreement with theory and verify the effectiveness of microwave frequency converters based on coherent optical micro-combs, with the ability to achieve reduced size, complexity, and potential cost.**

*Index Terms*—Microwave photonics, microwave frequency converters, Kerr optical comb.

## I. INTRODUCTION

Microwave frequency conversion for signal transmission or processing is a key processing block in radio-over-fibre (RoF) systems, beamforming and radio frequency communication networks [1-4]. As compared with electrical approaches that are subjected to the electrical bandwidth bottleneck [5, 6], photonic microwave frequency converters [7-9] could offer many competitive advantages including large bandwidth, high isolation, and strong immunity to electromagnetic interference, and so are promising solutions to meet with the ever-increasing demands for improved processing speed and performance.

Photonic microwave frequency converters are generally achieved by modulating an input microwave or radio frequency (RF) signal (with angular frequency $\omega_{RF}$) and a local oscillator (LO) signal (with angular frequency $\omega_{LO}$) onto an optical carrier and beat them upon photo-detection to produce the target intermediate frequency (IF) signal (with an angular frequency of $\omega_{IF} = \omega_{LO} \pm \omega_{RF}$). Many approaches to implement photonic microwave frequency converters have been reported, including those based on cascaded or parallel intensity modulators [7-12] and phase modulators [13]. Although diverse functions have been demonstrated for these frequency converters, they face limitations brought about by the external electrical LO sources, which suffer from the significantly increased cost and size for multi-stage frequency multiplication and the greatly degraded spectral purity at high frequencies, thus facing huge challenges to operate at high frequencies >40 GHz. Optoelectronic oscillators can address these limitations [14, 15], but are still limited by the limited operational bandwidth caused by the electrical components (e.g., electrical amplifiers and narrow-band filters) and bulky system size involving fibre spools [16], for example.

Kerr optical micro-combs [17-24], particularly those in CMOS-compatible platforms [25-30], can offer many distinctive advantages to perform as an equivalent LO source for photonic microwave frequency converters. This includes the ability to generate high-frequency electrical signals ranging from 10 GHz up to 500 GHz (determined by the comb spacing), a high spectral purity enabled by the ultra-high coherence of the generated states (e.g., Turing patterns, solitons, soliton crystals etc.), and a greatly reduced footprint (chip-scale) and complexity.

Here, we report a broadband LO-free photonic microwave frequency converter based on an integrated Kerr micro-comb source. Coherent micro-combs with a 48.9-GHz free spectral range (FSR) are generated by a high-Q micro-ring resonator (MRR) and serve as an equivalent millimeter-wave LO for microwave frequency conversion, thus eliminating the need of electronic LO sources and enabling wideband microwave frequency conversion within a greatly reduced potential footprint and cost. Meanwhile the optical parametric oscillation enables pure LO generation with a high frequency up to 48.9-GHz and even higher by utilizing multiple-FSR-spaced comb lines, thus enabling ultra-wideband microwave frequency conversion up to the U-band (40 to 60 GHz), which is extremely challenging for traditional photonic or electrical frequency convertors. In our experiments, the performance of the photonic microwave frequency converter is characterized at RF frequencies up to 40 GHz, achieving a ratio of −6.8 dB between output RF power and IF power, and a spurious suppression ratio of > 43.5 dB. This microcomb based photonic microwave frequency converter, with an ultra-high frequency photonic LO and greatly reduced system size, complexity, and potential cost, is a promising candidate for frequency conversion in modern radar and RoF systems.

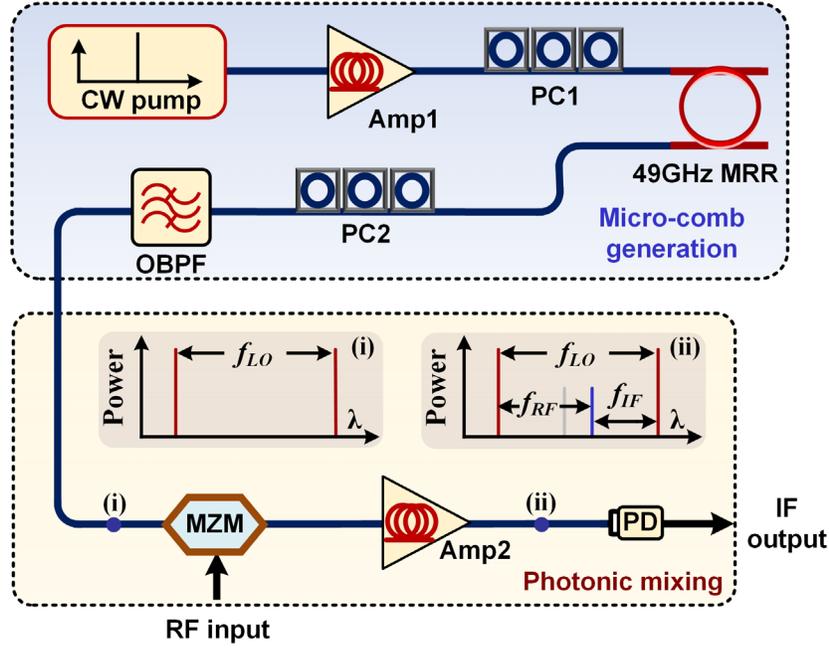

Fig. 1. Schematic diagram of the LO-free photonic microwave frequency converter based on an integrated optical micro-comb source. Amp: erbium-doped fibre amplifier. PC: polarization controller. MRR: micro-ring resonator. OBPF: optical bandpass filter. MZM: Mach-Zehnder modulator. PD: photodetector.

## II. PRINCIPLE

Figure 1 shows a schematic diagram of the LO-free photonic microwave frequency converter. Coherent optical frequency combs were generated in an integrated MRR, which was pumped by a continuous-wave (CW) laser amplified by an erbium-doped fibre amplifier, with the polarization adjusted via a polarization controller to optimize the power coupled into the MRR. When the pump wavelength was swept across one of the MRR resonances with the pump power high enough to provide sufficient parametric gain, optical parametric oscillation occurred, ultimately generating Kerr optical combs with a spacing equal to the FSR of the MRR (~48.9 GHz). By properly setting the pump power and wavelength detuning, we succeeded in operating the Kerr optical micro-comb in a high coherence state, with a spectral output consistent with operation in the soliton crystal regime [24] that exhibits high coherence and low RF intensity noise. The low-noise micro-comb served as a high-performance millimeter-wave LO source for microwave frequency conversion, with the added benefits of significantly reduced complexity, device footprint, and potential cost.

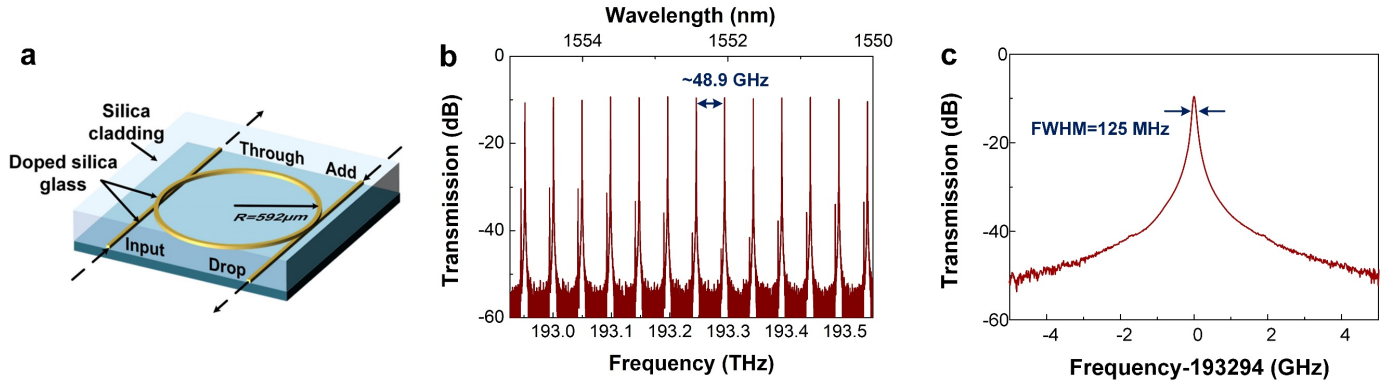

Fig. 2. (a) Schematic illustration of the 48.9 GHz-FSR MRR. (b) Drop-port transmission spectrum of the integrated MRR with a span of 5 nm, showing an FSR of 48.9 GHz, and (c) a resonance at 193.294 THz with a full-width at half-maximum (FWHM) of ~125 MHz, corresponding to a Q factor of ~1.55×10^6.

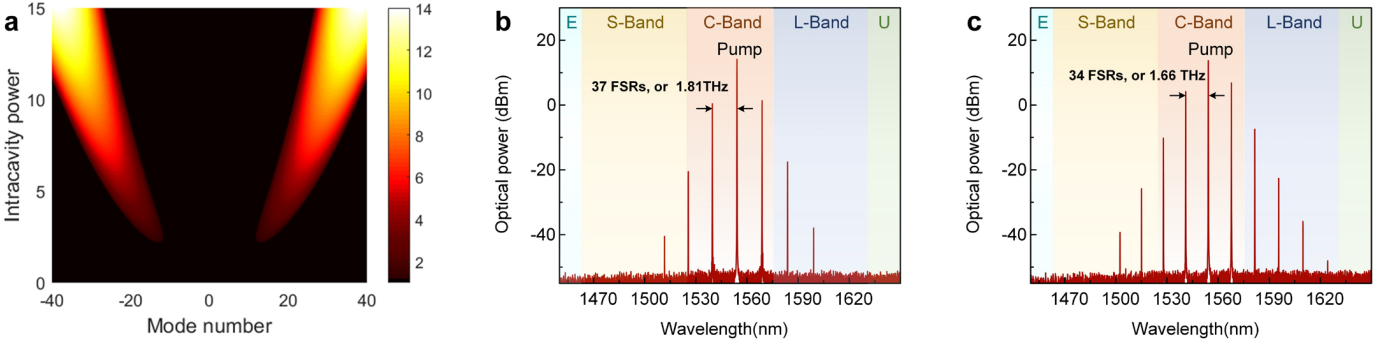

Fig. 3. (a) Simulated parametric gain with linear scales and normalized units. Measured optical spectra of the primary combs with spacings of (b) 1.81 THz and (c) 1.66 THz.

Moreover, the 48.9 GHz comb spacing also enabled an ultra-broad bandwidth for photonic microwave frequency conversion from the L-band to the U-band.

An optical bandpass filter was employed to select two adjacent comb lines from the coherent Kerr comb source for photonic microwave frequency conversion, and the optical field can be written as

$$u_{LO} = E_0[e^{j\omega t} + e^{j(\omega + \omega_{LO})t}] \tag{1}$$

where $E_0$ is the amplitude of the selected optical comb lines, $\omega$ and $(\omega + \omega_{LO})$ are the angular frequencies, with $\omega_{LO}$ denoting the angular frequency interval between them. The two comb lines were then modulated by an RF input ($V_{RF} \cdot \cos\omega_{RF}t$) via a Mach-Zehnder modulator, to produce an optical field given by

$$u_{MZM} = \frac{1}{2}E_0[e^{j\omega t} + e^{j(\omega + \omega_{LO})t}] \cdot$$
$$[e^{j(\varphi + \gamma/2 \cdot \cos\omega_{RF}t)} + e^{-j\gamma/2 \cdot \cos\omega_{RF}t}] \tag{2}$$

where $\varphi = \pi V_{dc} / V_\pi$ is the phase shift induced by the DC bias voltage, $\gamma = \pi V_{RF} / V_\pi$ is the modulation index, with $V_{dc}$ and $V_\pi$ denoting the DC voltage and the half-wave voltage, respectively.

Finally, the optical signals were amplified and converted back into electrical domain by a photodetector. With the Mach-Zehnder modulator biased at quadrature, the output electrical signal can be expressed as

$$\begin{aligned}
I_{out} &= u_{MZM} \cdot u_{MZM}^* \\
&= E_0^2\{[1 + \cos\varphi \cdot J_0(\gamma)] \\
&\quad + [1 + \cos\varphi \cdot J_0(\gamma)] \cdot \cos\omega_{LO}t \\
&\quad - 2J_1(\gamma) \cdot \sin\varphi \cdot \cos\omega_{RF}t \\
&\quad - 2J_2(\gamma) \cdot \cos\varphi \cdot \cos 2\omega_{RF}t \\
&\quad - J_2(\gamma) \cdot \cos\varphi \cdot \cos(\omega_{LO} + 2\omega_{RF})t \\
&\quad - J_2(\gamma) \cdot \cos\varphi \cdot \cos(\omega_{LO} - 2\omega_{RF})t \\
&\quad - J_1(\gamma) \cdot \sin\varphi \cdot \cos(\omega_{LO} + \omega_{RF})t \\
&\quad - J_1(\gamma) \cdot \sin\varphi \cdot \cos(\omega_{LO} - \omega_{RF})t\}
\end{aligned} \tag{3}$$

where $J$ denotes the Bessel function of the first kind. As reflected by the above equation, photonic microwave frequency conversion can be achieved in this manner, with the angular frequency of the output IF signal given by $(\omega_{RF} + \omega_{LO})$ or $(\omega_{LO} - \omega_{RF})$. We note that spurious frequencies are also generated, including $(\omega_{LO} - 2\omega_{RF}) = (2\omega_{IF} - \omega_{LO})$ that appeared in our experiments.

## III. EXPERIMENT

The MRR used to generate the Kerr optical comb (Fig. 2(a)) was fabricated on a high-index doped silica glass platform using CMOS-compatible fabrication processes [31-38]. First, high-index (n = ~1.7 at 1550 nm) doped silica glass films were deposited using plasma enhanced chemical vapour deposition, then patterned by deep ultraviolet photolithography and etched via reactive ion etching to form waveguides with exceptionally low surface roughness. Finally, silica (n = ~1.44 at 1550 nm) was deposited as an upper cladding. The advantages of our platform for optical micro-comb generation include ultra-low linear loss (~0.06 dB·cm⁻¹), a moderate nonlinear parameter (~233 W$^{-1}$·km$^{-1}$), and in particular negligible nonlinear loss up to extremely high intensities (~25 GW·cm$^{-2}$). The radius of the MRR was ~ 592 μm, corresponding to an FSR of ~0.4 nm or~48.9 GHz (Fig. 2(b)). The relatively small FSR of the MRR led to a comb spacing falling in the millimeter-wave region, bridging the gap between the frequency scales of on-chip optical micro-combs and microwave systems. The ultra-low loss of the MRR resulted in a Q factor of ~1.5 million (Fig. 2(c)). After packaging the device with fibre pigtails, the through-port insertion loss was ~1 dB, assisted by on-chip mode converters.

To generate coherent micro-combs, the CW pump power was amplified to ~30.5 dBm and the wavelength swept from blue to red. When the detuning between the pump wavelength and MRR's cold resonance wavelength became small enough such that the intracavity power reached a threshold, modulation instability driven oscillation was initiated [25]. Primary combs were generated with a spacing of multiple FSRs, determined mainly by the intra-cavity power and dispersion. The calculated parametric gain (Fig. 3(a)) could be controlled by varying the pump wavelength detuning, which in turn resulted in different spacings of the generated primary combs (Fig. 3(b), (c)) [39, 40].

As the detuning was changed further, distinctive 'fingerprint' optical spectra were observed (Fig. 4) which were similar to the spectral interference between tightly packed solitons in the cavity – so called "soliton crystals" that have been reported [24]. The soliton crystal step in the measured transmission (Fig. 4(c)) and the dramatic reduction of the RF intensity noise (Fig. 4(d)) are hallmarks of coherent micro-combs. Therefore, the ~ 48.9 GHz-spaced coherent combs were able to serve as an equivalent LO source, reaching the millimetre-wave region (U band, 40 to 60 GHz) – a regime that is challenging to achieve using traditional electrical methods [15]. Moreover, the optical power of the two comb lines near ~1541 nm (selected for the photonic microwave frequency conversion in our experiment) reached over 2 dBm (Fig. 4(a), zoom-in view), and the power ratio of the pump to the comb lines was -11 dBm (Fig. 4(b)), which provided a relatively high link gain for the microcomb-based microwave signal processors [41-43].

By changing the pump power within ± 0.5 dB, a diverse range of spectra were observed, all displaying low RF noise, and all being indicative with different soliton crystal superstructures [44] (Fig. 4, Fig. 5). This not only enabled the generation of coherent micro-combs for photonic microwave frequency conversion, but also offered a range of different comb spectral shapes suited to different applications. The formation of high quality stable combs, potentially consisting of soliton crystals, was readily achievable using straightforward adiabatic pump wavelength sweeping. We found that it was not necessary to achieve a rigorous single soliton state in order to achieve high microwave frequency conversion performance — only that the chaotic regime needed to be avoided. This is important since there are a much wider range of coherent low RF noise states that are more readily accessible than just the single soliton state [31, 39]. We employed a TEC (Thermo-Electric Cooler) controller to stabilize the chip temperature and the generated soliton crystal states were stable for over 12 hours without stabilization (other than the TEC) or feedback.

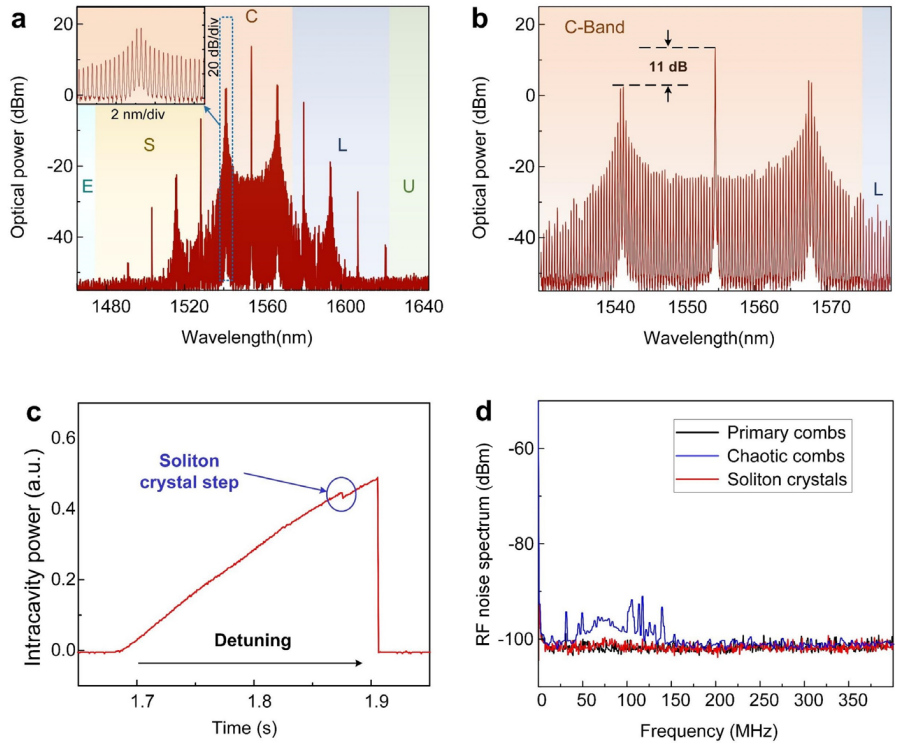

Fig. 4. Measured optical spectrum of the generated micro-comb with a span of (a) 200 nm and (b) 50 nm. (c) Transmission of the pumping resonance and (d) the RF spectra corresponding to different regimes of micro-comb dynamics.

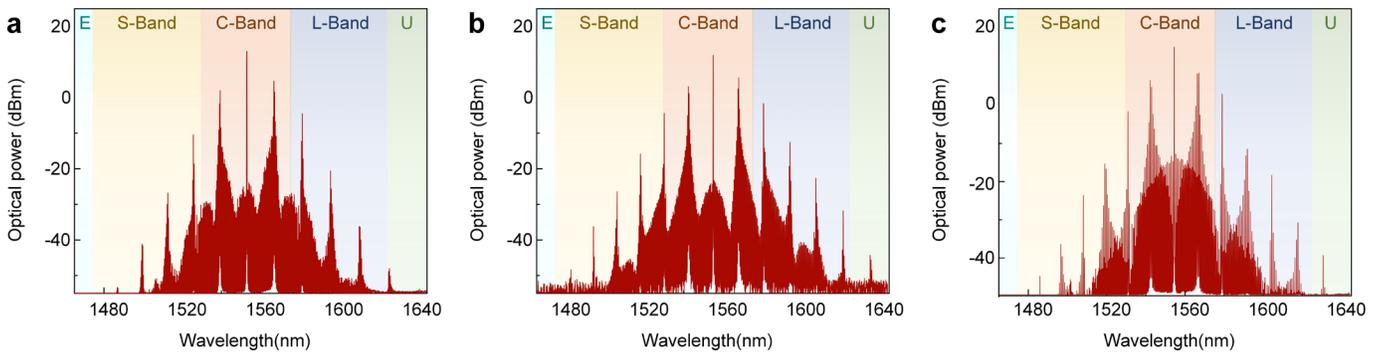

Fig. 5. Optical spectra of the generated micro-combs showing different superstructures reminiscent of soliton crystals.

To generate the equivalent photonic LO, two comb lines near ~1541 nm were filtered and modulated by RF signals (with 14 dBm power) with different frequencies ranging up to 40 GHz. The corresponding optical spectra are shown in Fig. 6 (a). Next, the optical signals were amplified to 12 dBm and converted back into electrical domain where the frequency mixing of the RF signal and the LO signal was achieved via photo-detection. Figure 6(b) shows the measured electrical spectral output of the IF signal. As the input RF frequency $f_{RF}$ was varied from 40 GHz to 23 GHz (from the Ka-band to the X-band), the converted IF output ($f_{IF} = f_{LO} - f_{RF}$) varied from 8.9 GHz to 25.9 GHz (from the X-band to the Ka-band) with a power variation of < 5 dB, reflecting the broad

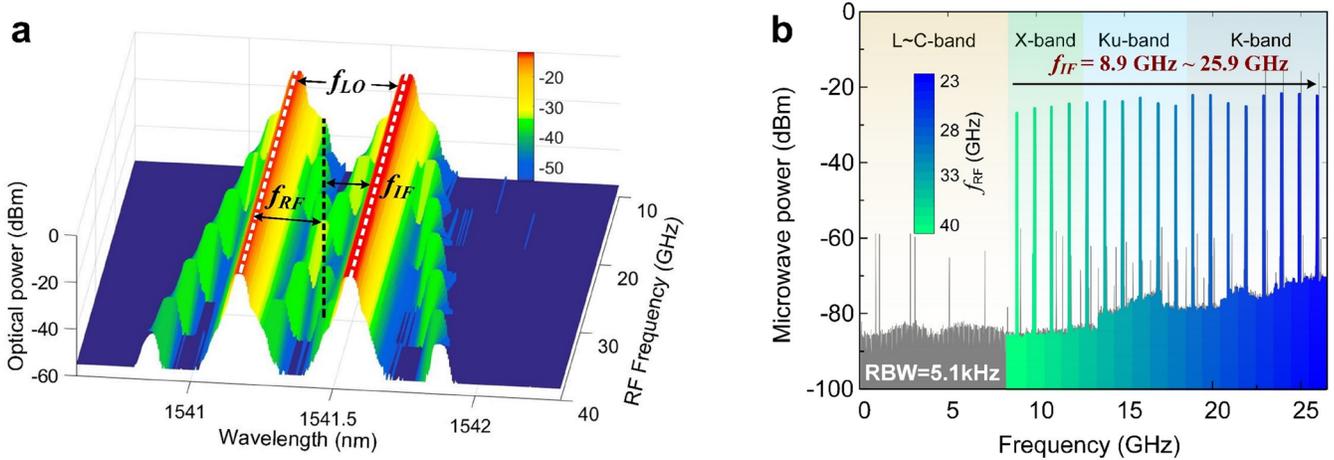

Fig. 6. Measured (a) optical spectra of the equivalent photonic LO and (b) output IF electrical spectra, with the input RF frequency $f_{RF}$ varying from 40 GHz to 23 GHz.

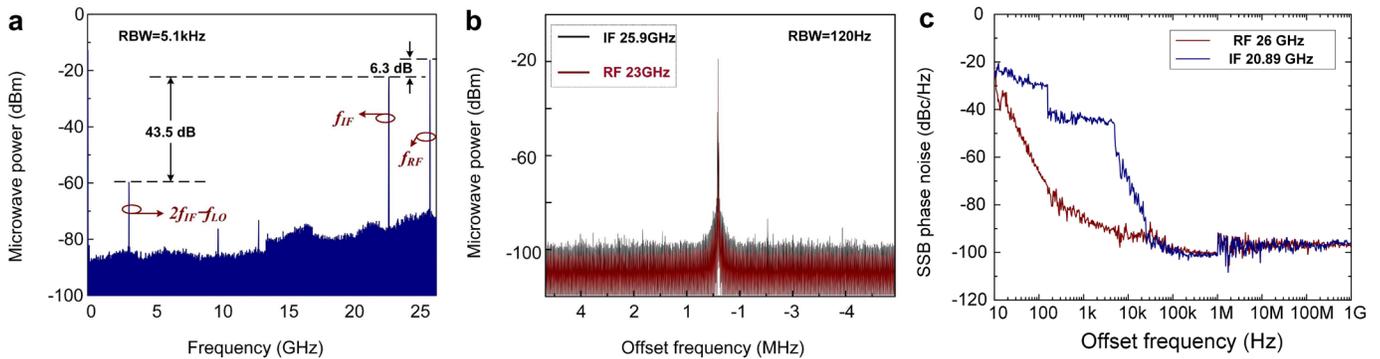

Fig. 7. Measured (a) output electrical spectra of the photonic microwave frequency converter with $f_{RF}$ = 26 GHz, (b) electrical spectra of the input RF signal and output IF signal in a span of 10 MHz, and (c) phase noise spectra of the input RF signal at 26 GHz and the IF signal at 20.89 GHz.

operational bandwidth of our photonic microwave frequency converter.

Figure 7(a) shows the output electrical spectra when $f_{RF}$ = 26 GHz, corresponding to an IF output of 22.9 GHz. The electrical spectra in a 10-MHz span (Fig. 7(b)) exhibited a signal-to-noise ratio of > 70 dB, which further confirmed the low noise performance of the photonic microwave frequency converter. The output power ratio of the IF to RF signals reached −6.3 dB, while the conversion efficiency of the output IF relative to the input RF power was −36.4 dB, with the −30.3dB link gain included. The spurious suppression ratio (power ratio of the IF signal to the spurious signal at $2f_{IF} − f_{LO}$) was 43.5 dB.

The phase noise performance of the input RF signal at 26 GHz and an output IF signal at 20.89 GHz was also measured using an RF spectrum analyzer (Keysight N9010A), as shown in Fig. 7(c). Here the 20.89 GHz IF tone was the converted output of a 28 GHz RF tone, instead of the measured 26 GHz RF tone, for the phase noise measurement. Yet since a 2 GHz difference in the RF tone's frequency would not significantly vary its phase noise performance, the phase noise spectra measured here could still reflect the performance of our photonic LO. The phase noise spectrum of the IF signal shows low residue noise brought about by the photonic LO at offset frequencies from 20 kHz to 1 GHz, verifying the microcomb's low noise performance that is promising for ultra-high frequency photonic LO generation and thus wideband microwave frequency conversion.

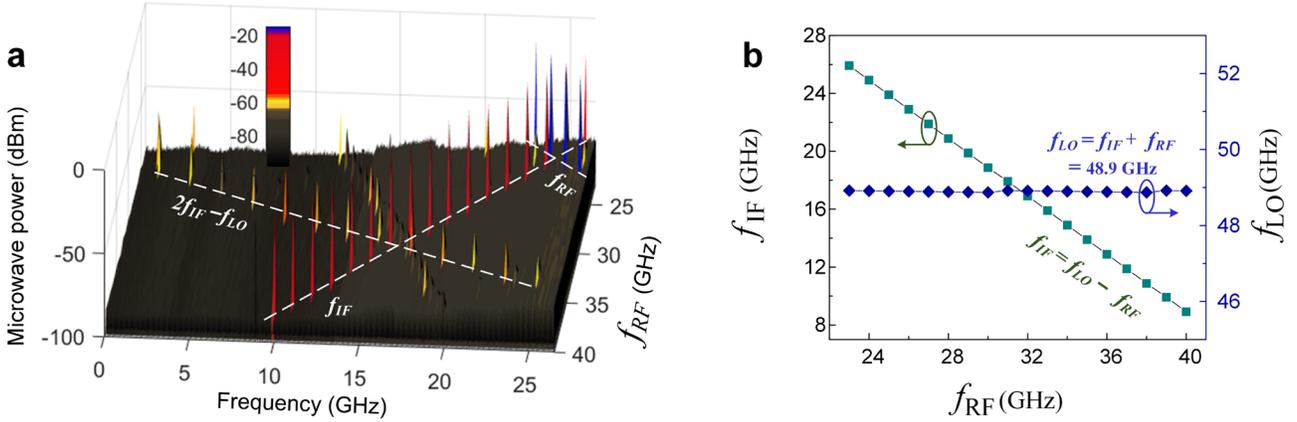

Fig. 8. (a) Measured output electrical spectra of the photonic microwave frequency converter with input RF frequency $f_{RF}$ varying from 40 GHz to 23 GHz. (b) Extracted $f_{IF}$ and calculated $f_{LO} = f_{IF} + f_{RF}$ = 48.9 GHz.

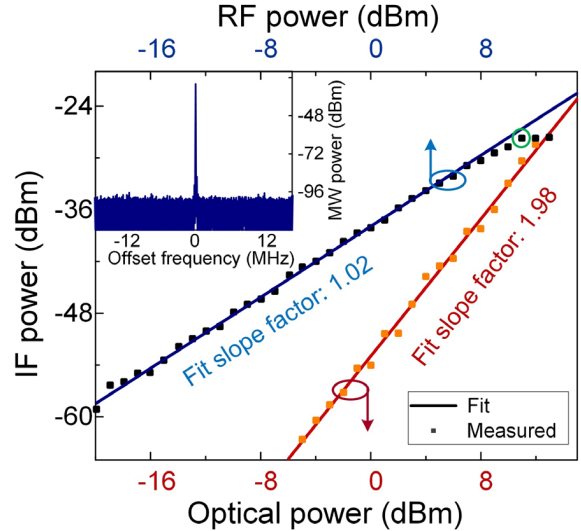

Fig. 9. Two stacked plots of measured IF power for different input RF power and received optical power at the photodetector, respectively. The RF and optical power was fixed at 14 dBm and 12 dBm, respectively, when varying the other.

We note that the deterioration in the IF phase noise for offset frequencies < 20 kHz can be optimized by stabilizing the pump power [45] and setting the detuning to the quiet operation point [46]. Further optimization would need to deal with other noise sources such as the relative intensity noise of the pump and the thermal refractive noise of the resonator [47].

The frequency relationship between $f_{RF}$ and the measured $f_{IF}$ is clearly shown in Fig. 8, together with the calculated $f_{LO} = f_{IF} + f_{RF}$ = 48.9 GHz. This verifies the generation of a millimeter-wave LO using our approach and indicates the potential for ultra-wideband microwave frequency conversion from the L- to U-bands (i.e., 0 to 60 GHz). Note that our approach is capable of achieving even higher frequencies through the use of multiple-FSR-spaced comb lines. However, we were experimentally limited in our measurement capability by the bandwidth of our electrical spectrum analyzer and the RF source. As shown in Fig. 9, the IF power had a linear dependence on the input RF power (with a slope of 1.02, saturating at 10 dBm) and the optical power received at the photodetector (with a slope of 1.98), which matches closely with the theoretical predictions of 1 and 2, respectively from Eq. 3.

For applications requiring further suppression of spurious frequencies (such as $2f_{\text{IF}} - f_{\text{LO}}$), the two comb lines can be separated into two paths, where one is modulated by a carrier-suppressed double-sideband format to generate photonic RF sidebands, and the other serves as the photonic LO sideband for frequency mixing. In addition, though the LO frequency was not tunable in this paper, recent advances in dual micro-comb generation [48-49] provide possibilities to achieve a frequency-tunable photonic LO for the microwave frequency converter. By changing the relative offsets of the two combs via thermal tuning, an almost unlimited range of photonic LO frequencies can be realized from sub-GHz to beyond even a terahertz.

## IV. CONCLUSION

We demonstrate a broadband LO-free photonic microwave frequency converter based on an integrated Kerr comb source. Coherent micro-combs with a 48.9-GHz spacing were generated by a high-Q MRR and served as an equivalent millimeter-wave LO for microwave frequency conversion, enabling microwave frequency conversion up to the U-band. We experimentally verified the photonic microwave frequency converter's performance for input RF frequencies up to 40 GHz, achieving a ratio of −6.8 dB between output RF power and IF power and a spurious suppression ratio of > 43.5 dB. With an ultra-high frequency photonic LO and greatly reduced system size, complexity, and potential cost, this microcomb based photonic microwave frequency converter is attractive for frequency conversion in radar and RoF systems.

## V. ACKNOWLEDGEMENTS

This work was supported by the Australian Research Council Discovery Projects Program (No. DP150104327). RM acknowledges support by the Natural Sciences and Engineering Research Council of Canada (NSERC) through the Strategic, Discovery and Acceleration Grants Schemes, by the MESI PSR-SIIRI Initiative in Quebec, and by the Canada Research Chair Program. He also acknowledges additional support by the Government of the Russian Federation through the ITMO Fellowship and Professorship Program (grant 074-U 01) and by the 1000 Talents Sichuan Program in China. Brent E. Little was supported by the Strategic Priority Research Program of the Chinese Academy of Sciences, Grant No. XDB24030000.